\documentclass[aps,showpacs,twocolumn,preprintnumbers,amsmath,amssymb]{revtex4}

\usepackage{amsmath,amssymb}
\usepackage{graphics,epsfig}
\usepackage{graphicx}
\usepackage{color}

\newcommand{\be}{\begin {equation}}
\newcommand{\ee}{\end {equation}}
\newcommand{\bea}{\begin {eqnarray}}
\newcommand{\eea}{\end {eqnarray}}
\newcommand{\ovl}{\overline}

 \newcommand{\vct}[1]{{\mbox {\boldmath $#1$}}}
\def\build#1_#2^#3{\mathrel{\mathop{\kern 0pt#1}\limits_{#2}^{#3}}}

\begin{document}

\title{Matrix Exponential-Based Closures for the Turbulent Subgrid-Scale Stress Tensor}

\author{Yi Li$^{1,2}$, Laurent Chevillard$^{1,3}$,  Gregory Eyink$^4$ and Charles Meneveau$^1$}
\affiliation{$^1$Department of Mechanical Engineering and Center
of Environmental and Applied Fluid Mechanics, The Johns Hopkins
University, Baltimore, MD, 21218\\ $^2$Department of Applied Mathematics, 
The University of Sheffield H23B, Hicks Building, Hounsfield Road
Sheffield, S3 7RH, UK\\ $^3$Laboratoire de Physique de l'\'Ecole Normale Sup\'erieure
de Lyon, CNRS, Universit\'e de Lyon, 46 all\'ee d'Italie F-69007 Lyon,
France\\$^4$Department of Applied
Mathematics \& Statistics, and Center
of Environmental and Applied Fluid Mechanics, The Johns Hopkins University,
Baltimore, MD, 21218}

\begin{abstract}

Two approaches for closing the turbulence subgrid-scale stress tensor in terms of
matrix exponentials are introduced and compared. The first approach
is based on a formal solution of the stress transport equation in which the
production terms can be integrated exactly in terms of matrix
exponentials. This formal solution of
the subgrid-scale stress transport equation is shown to be useful to explore
special cases, such as the  response to constant velocity
gradient,  but neglecting pressure-strain correlations and diffusion effects. The second approach is based on an Eulerian-Lagrangian change of variables, combined
with the assumption of isotropy for the conditionally averaged
Lagrangian velocity gradient tensor and with the `Recent Fluid
Deformation' (RFD) approximation. It is shown that both
approaches lead to the same basic closure in which the stress tensor is expressed as the
product of the matrix exponential of the resolved velocity gradient tensor multiplied by its transpose.  
Short-time expansions of the matrix exponentials are shown to
provide an eddy-viscosity term and particular quadratic terms, and
thus allow a reinterpretation of traditional eddy-viscosity and nonlinear stress closures.
The basic feasibility of the matrix-exponential closure is
illustrated by implementing it successfully  in Large Eddy
Simulation of forced isotropic turbulence. The matrix-exponential 
closure employs the drastic approximation of entirely omitting the pressure-strain correlation and 
other `nonlinear scrambling' terms. But unlike eddy-viscosity closures,
the matrix exponential approach provides a simple and local closure
that can be derived directly from the stress transport equation with the production term, 
and using physically motivated  assumptions about  Lagrangian  decorrelation and  upstream isotropy. 

 \end{abstract}

\maketitle

\section{Introduction}
One of the most basic challenges in turbulence modeling is the
need for closures for the fluxes associated with unresolved
turbulent fluctuations.  In the context of Large Eddy Simulation
(LES), closures are required for the subgrid-scale (SGS) stress
tensor \cite{LesMet96,MenKat00}. Traditional closures involve
mostly algebraic expressions relating the stress tensor to powers
of the velocity gradient tensor. More elaborate approaches using
separate transport equations have sometimes also
been employed, although these tend to be significantly more costly
in the context of LES. Closures expressing
the stress in terms of the matrix exponential function do not
appear to have received much attention in the literature. The
objective of the present work is to identify and discuss two
separate paths that lead to such closures. Both paths are based on
the Lagrangian dynamics of turbulence, i.e. on an understanding of
the evolution of turbulence as one follows fluid-particle paths in
time.

The use of Lagrangian concepts in turbulent flows has a long
history \cite{Kra65,TenLum72} and, in recent years, has seen
renewed interest for modeling \cite{Gir94,ChePum99,JeoGir03,PumShr03,LiMen06}.
Among others, a new model for the pressure-Hessian tensor
based on the recent Lagrangian evolution of
fluid elements  -- the Recent Fluid Deformation (RFD) closure --
has been proposed \cite{CheMen06,CheMen07,CheMen08}. In this approach, a
change of variables is made expressing spatial gradients in terms
of Lagrangian gradients (e.g. how does a variable at the present
location vary if we change the initial position of the fluid
particle at an earlier time). Then the assumption of isotropy is
introduced for the Lagrangian gradient tensors.  This assumption allows simpler isotropic 
forms to be used, and is argued to be justified based on Lagrangian decorrelation effects. Deviations from
isotropy at the present location for the Eulerian gradient tensors
develop as a result of fluid material deformation along the
Lagrangian trajectory. More traditionally, the Lagrangian time evolution of the 
stress tensor following fluid particles can be derived by taking 
appropriate moments of the Navier-Stokes equations. 
In this paper we examine both of these approaches to
formulate models for the SGS stress tensor in turbulence 
in the context of LES. 

A description of small-scale structure of turbulence begins with the Navier-Stokes equations of an incompressible fluid of velocity $\textbf{u}$:
\begin{equation}\label{eq:NS}
\frac{d\textbf{u}}{dt} = \frac{\partial\textbf{u}}{\partial t}+(\textbf{u}\cdot
\vct{\nabla})\textbf{u}=-\vct{\nabla}p+\nu\vct{\nabla}^2\textbf{u}\mbox{ ,}
\end{equation}
where $d/dt$ stands for the Lagrangian material derivative, $p$
the pressure divided by the density of the fluid and $\nu$ the
kinematic viscosity.  Because of incompressibility, the velocity gradient tensor $A_{ij}=\partial u_i/\partial x_j$
must remain trace-free, i.e. $A_{ii}=0$, and the pressure
field is the solution of the Poisson equation
$\nabla^2
p=-A_{lk}A_{kl}$.

In the framework of LES, the SGS stress tensor is defined
using the filtering approach \cite{Leo74,Ger92}, 
\be \label{eq:DefSubgrid}
\tau_{ij} = \overline{u_iu_j} -
\overline{u}_i\overline{u}_j \mbox{ .}
\ee 
An overbar denotes spatial
filtering at a scale $\Delta$ and 
is formally given by a convolution with a nonnegative, 
spatially well-localized filtering function $\mathcal G(\textbf{r})$ of characteristic size $\Delta$, 
with unit integral $\int \mathcal G(\textbf{r}) d\textbf{r}= 1$, 
namely
$$\overline{\textbf{u}} (\textbf{x},t)= \int 
\mathcal G(\textbf{r})\textbf{u} (\textbf{x+\textbf{r}},t) d\textbf{r}\mbox{ .}$$

The SGS tensor $\vct{\tau}$ enters in the dynamics of 
the filtered velocity $\overline{\textbf{u}}$ as it can be seen 
when applying the filtering procedure to the Navier-Stokes equations (Eq. (\ref{eq:NS})), 
\begin{equation}\label{eq:FNS}
\frac{D\overline{\textbf{u}}}{Dt} = \frac{\partial\overline{\textbf{u}}}{\partial t}+(\overline{\textbf{u}}\cdot 
\vct{\nabla})\overline{\textbf{u}}=-\vct{\nabla}\overline{p}+\nu\vct{\nabla}^2\overline{\textbf{u}} 
- \vct{\nabla} \cdot \vct{\tau}\mbox{ ,}
\end{equation}
where $D/Dt$ stands for the Lagrangian material derivative with 
$\overline{\textbf{u}}$ as the advecting velocity, and $\overline{p}$
the filtered pressure divided by the density of the fluid. 
Because of incompressibility, the filtered 
velocity gradient tensor $\overline{A}_{ij}=\partial \overline{u}_i/\partial x_j$
must remain trace-free, i.e. $\overline{A}_{ii}=0$, and the filtered pressure
field is the solution of the respective Poisson equation
$\nabla^2
\overline{p}=-\overline{A}_{lk}\overline{A}_{kl}-\partial^2\tau_{ij}/\partial x_i\partial x_j$.

Next, we also consider the transport equation for the SGS stress
tensor $\vct{\tau}$ \cite{Leo74,Dea74,Ger92}, which follows from Eq. (\ref{eq:FNS}):
 \be
\label{eq:TranspEquTau} \frac{D\vct{\tau}}{Dt} =\frac{\partial
\vct{\tau}}{\partial
t}+(\overline{\textbf{u}}\cdot\vct{\nabla})\vct{\tau}=-\vct{\tau}
\overline{\textbf{A}}^\top-
\overline{\textbf{A}}\vct{\tau}+\vct{\Phi} \mbox{ ,} 
\ee 
where the
term $\vct{\Phi} \equiv \vct{\Phi}_p+\vct{\Phi}_\nu -
\vct{\nabla}\cdot \vct{\mathcal J}$ includes the pressure gradient-velocity correlation
$$\Phi_{p,ij} = -\left[ \overline{u_i\partial_jp}-\overline{u_i}
\partial_j\overline{p}+ \overline{u_j\partial_ip}- \overline{u_j}
\partial_i\overline{p} \right]\mbox{ ,}$$
the viscous term, 
$$ \Phi_{\nu,ij} = \nu(
\overline{u_i\vct{\nabla}^2u_j}-\overline{u}_i
\vct{\nabla}^2\overline{u}_j+\overline{u_j\vct{\nabla}^2u_i}
-\overline{u}_j\vct{\nabla}^2\overline{u}_i)\mbox{ ,}$$
and the generalized central third-order moment
$$\mathcal J_{ijk}= \overline{u_iu_ju_k}-\overline{u}_j\tau_{ik}-
\overline{u}_i\tau_{jk}-\overline{u}_k\tau_{ij}- \overline{u_i}\
\overline{u_j} \overline{u}_k\mbox{ .}$$

In \S \ref{sec_stressequation} it is shown that a formal solution for
the stress transport equation may be obtained by integrating the
production term exactly. This solution, suggested by \cite{Crow68}
but---to our knowledge---little pursued, will  be shown to
involve matrix exponentials.  The developments presented require some 
assumptions of Lagrangian isotropy and 
decorrelation, and some empirical evidence supporting these assumptions is provided in \S 
\ref{sec_dataDNS} based on results from Direct Numerical Simulations (DNS). 
In \S \ref{sec_LFDclosure} the RFD closure for the SGS stress is developed. 
The resulting model is shown to be expressible compactly in terms of matrix 
exponentials as well.  Differences and similarities
between the RFD and transport equation solutions are discussed.
 In \S \ref{sec_expansions}, the matrix-exponential 
solutions are expanded for short times. The expansions allow to establish
relationships to traditional eddy-viscosity and nonlinear closure models in
turbulence. In \S \ref{sec_testsinLES} the matrix-exponential
closure is implemented in a most simple flow to illustrate its feasibility and cost.

\section{Solution to stress transport equation using matrix exponentials}
\label{sec_stressequation}

Equation \ref{eq:TranspEquTau}  is of the form of the
``time-dependent Lyapunov equation'', if the tensor $\vct{\Phi}$'s implicit dependencies
upon the velocity fluctuations and the stress tensor were disregarded (in reality, $\vct{\Phi}$ depends upon small-scale velocity fluctuations and thus the full equation is highly non-linear and non-local).
The formal solution of the Lyapunov equation
in terms of matrix exponentials has been found useful in a number of
other fields: principal oscillation pattern analysis
\cite{Penland89},  mechanics of finite deformations \cite{TruNol92},
and fluctuation-dissipation theorems
for stochastic linear systems \cite{Eyink98}. In the context of
the SGS stress transport equation the solution at time $t$
(starting from an initial condition at time $t'$) may be written
formally as follows
 \begin{equation}\label{eq:ExactSol}
\vct{\tau}(t) = \textbf{H}(t,t') \vct{\tau}(t')\textbf{H}^\top(t,t') +
\int_{t'}^t \textbf{H}(t,s)
\vct{\Phi}(s)  \textbf{H}^\top(t,s) ds  \mbox{ ,}
\end{equation}
where
\begin{equation}\label{eq:ODEforE}
\frac{D\textbf{H}(t,t')}{Dt}=-\overline{\textbf{A}}(t)\textbf{H}(t,t')
\mbox{ and     } \textbf{H}(t',t')={\bf I} \mbox{ .}
\end{equation}
To our knowledge, this approach to solve the stress equation in RANS 
closures was first suggested in the turbulence literature by \cite{Crow68} 
(see equation (4.4) in this reference). For the general case of time-varying velocity
gradient,  we note that the auxiliary matrix $\textbf{H}(t,t')$ can be
written as a time-ordered exponential (see Refs. \cite{DolFri79,Eyink98,Rugh96} for background on
this basic matrix function)
$$\textbf{H}(t,t') = {\rm Texp}^{^+}\left[-\int_{t'}^t
\overline{\textbf{A}}(s)\,ds\right]\mbox{ .}$$

Equation (\ref{eq:ExactSol}) illustrates clearly the distinct roles
played by the production term and the contribution given
by $\vct{\Phi}$. Evaluation of  Eq. \ref{eq:ExactSol}  requires the knowledge of
the time history of $\overline{\textbf{A}}(s)$ as well as accurate
closures for $\vct{\Phi}(s)$ along the fluid history $t' < s <t$.

As a next step, one may consider the special case in which the
velocity gradient is considered to be constant between the initial
time $t'$ and $t$, and set equal to (e.g.)
$\overline{\textbf{A}}(t)$ and simply denoted by
$\overline{\textbf{A}}$. For this approximate situation, the
solution of Eq.(\ref{eq:ODEforE}) may be written as an ordinary
matrix exponential $\textbf{H}(t,t') = e^{-(t-t'){\overline{\bf
A}}}$, where the matrix exponential is defined in the usual way
$e^{ {{\bf B}}} = \sum_{n=0}^{+\infty}{{\bf B}}^n/n!$.  
To simplify further, consider Eq. \ref{eq:ExactSol} for  the  case
$\vct{\Phi}=0$, i.e. now retaining only the production term. This step
eliminates the important isotropization effects
of pressure-strain and also the nonlinear diffusion effects of the transport
terms.  While clearly missing important physics, it is still instructive to
observe that this simplification allows the solution (Eq. \ref{eq:ExactSol}) to be written as: 
\begin{equation}\label{eq:NonCondStress}
\vct{\tau}(t) =
~e^{-(t-t')\overline{{\bf A}} }~{\vct{\tau}}(t')~
e^{-(t-t')\overline{{\bf A}}^\top }\mbox{ .}\end{equation}

At this stage it is conceptually advantageous to make connection
with Refs.  \cite{Adr90,LanMos99,Pop00}, where it is
proposed to use conditional statistics to capture the relevant
statistics of the SGS stress. For example, in Ref. \cite{LanMos99}, it is shown
that the least-square-error best estimate for the SGS stress is of
the form of a multi-point conditional average, namely $ \langle
\tau_{ij}~\vert ~\overline{{\bf u}}_1, \overline{{\bf
u}}_2,...,\overline{{\bf u}}_N\rangle$. The multi-point
conditioning variables $\{ \overline{{\bf u}}_1, \overline{{\bf
u}}_2,...,\overline{{\bf u}}_N\}$ are, in principle, constituted
by the {\it entire} ($N$-point) resolved velocity field at scale
$\Delta$. To simplify the conditioning, one may limit the
information to the past time-history of the local velocity
structure. In particular, a good choice that captures much of the
local dynamics in a Galilean invariant fashion is the Lagrangian
past history of the filtered velocity gradient tensor
${\overline{\bf A}}$. The
dependence on the Lagrangian time history along a fluid particle
advected by the filtered resolved velocity field is thus assumed
to be described by $\overline{{\bf A }}(s)$ with $s \leq t$ (here and below,
the dependence of $\overline{{\bf A }}$ on spatial position is omitted
for clarity). According to these ideas, we define a \textit{quasi-optimal} 
SGS stress tensor $\vct{\tau}^{(o)}(t)$ as the conditional average 
$\vct{\tau}^{(o)}(t)=\langle \vct{\tau}(t) \vert \overline{{\bf A}} \rangle$. 
Noticing that the matrix exponential prefactors entering in Eq. \ref{eq:NonCondStress} 
are some deterministic functions of the velocity gradient tensor itself, 
they thus can be taken out from this conditional average and we get the following  stress tensor:
\be \label{eq:stress-simple}
\vct{\tau}^{(o)}(t)  =
~e^{-(t-t')\overline{{\bf A}} }~\langle \vct{\tau}(t') \vert
\overline{{\bf A}} \rangle~ e^{-(t-t')\overline{{\bf A}}^\top
}\mbox{ .}
\end{equation}
With this expression, the closure problem has been changed from requiring a model for
the local stress tensor at time $t$ to requiring a model for the conditional average of the
`upstream initial condition' at time $t'<t$. The initial condition needed is a symmetric tensor. 
In the absence of additional information, the simplest assumption is to postulate that this 
conditionally averaged `upstream' stress tensor is isotropic, namely 
\be \langle  \tau_{ij}(t') \vert \overline{{\bf A}}\rangle ~ \approx 
~{1 \over 3} \langle  \tau_{kk}(t') \vert \overline{{\bf A}} \rangle ~\delta_{ij}.
\ee

The magnitude of the tensor is proportional to the trace of the SGS tensor and has units of squared velocity.  
The assumption of isotropy may be justified if $\vct{\tau}(t')$ 
and $\overline{{\bf A}}(t)$ become more and more de-correlated as the elapsed 
time $t-t'$ grows, then no locally strong and statistically preferred direction should exist. 
This step introduces a characteristic decorrelation time-scale $\tau_a$, and  
$t-t'$ will be chosen to be of the order of such a decorrelation time-scale.  
Clearly, one must also assume local isotropy to hold for the statistics, and this 
is justified from the usual arguments in turbulence when $\Delta$ is sufficiently 
small compared to the integral scale. Incidentally, it is expected that a 
decorrelation between $\vct{\tau}(t')$ and $\overline{{\bf A}}(t)$ may 
occur due to pressure effects,  turbulent diffusion, etc. Some numerical evidence for such decorrelation and isotropization  is provided in the next section.

The trace of the conditional SGS tensor, $\langle  \tau_{kk}(t') \vert \overline{{\bf A}}(t) \rangle$,  must still be specified. The simplest option that is consistent with a local evaluation of velocity and length-scales is to choose a factor proportional to $\Delta^2|\overline{\bf S}|^2$, where $\overline{\bf S}\equiv (\overline{\bf A} + \overline{\bf A}^\top)/2$
is the filtered strain rate tensor and $|\overline{\bf S}|\equiv  (2 \overline{S}_{ij}\overline{S}_{ij})^{1/2}$. 
Finally, replacing into
Eq. \ref{eq:stress-simple} with $t-t'=\tau_a$,  we get
 \be \vct{\tau}^{(o)} =  c_{\rm exp}
\Delta^2|\overline{\bf S}|^2 e^{-\tau_a\overline{\bf
A}}  e^{-\tau_a\overline{\bf A} ^\top } \mbox{ ,} \label{eq:stressTransportEq}
\ee
where the parameter $c_{\rm exp}$ is unknown and may be obtained by empirical knowledge, 
or by generalizing the dynamic model \cite{GerPio91}.

For completeness and clarity, we remark that the matrix exponential solution may
equivalently be obtained by solving the linearized equation for a
turbulent fluctuation that only keeps the
Rapid Distortion term from the large-scale velocity field, and neglects all other effects. 
That is to say, we solve formally the equation $D_t u_i^\prime = - u_k^\prime \overline{A}_{ik}$
using the matrix exponential function. The solution is then  multiplied by its transpose to form
$u_i^\prime(t) u_j^\prime(t)$ which is then  averaged over the fluctuating initial
condition $u_i^\prime(t') u_j^\prime(t')$ (conditioned on a constant $\overline{\textbf{A}}$).
The  averaging of the term $u_i^\prime(t') u_j^\prime(t')$
yields the initial (`upstream')  stress tensor $\vct{\tau}(t')$, and with the conditional averaging,  an expression equivalent to Eq. \ref{eq:stress-simple}  is obtained. This is similar to the equivalence between solving the equation for co-variances or for the fluctuations and then averaging, as noted in the context of stochastic linear systems in \cite{Eyink98}.

Equation (\ref{eq:stressTransportEq}) represents a closure for the SGS stress expressed in terms of matrix
exponentials instead of the more commonly used algebraic closures \cite{Pop75}.
In section \ref{sec_LFDclosure}, a connection is noted between the expression
Eq.  (\ref{eq:stressTransportEq}) and a physical closure for the subgrid stress tensor based on the recent fluid deformation closure in the Lagrangian frame.

\section{Empirical Evidence of Lagrangian decorrelation and Isotropy from DNS}
\label{sec_dataDNS}

In order to verify whether the decorrelation and isotropization of
conditional averages of SGS stresses occur in turbulence, we analyze
a DNS dataset of forced isotropic turbulence. The simulation is
conducted using a pseudo-spectral method in a $[0,2\pi)^3$ box.
$128^3$ grid points are used. Fourier modes in shells with $|{\bf
k}| < 2$ are forced by a term added to the Navier-Stokes
equations which provides constant energy injection rate $\epsilon_f
= 0.1$. The viscosity of the fluid is $\nu=0.0032$.  Data is
collected after the simulation reaches statistical steady state.
Note that $\vct{\tau}(t')$ is the SGS stress at a previous time $t'$
and at the spatial location occupied by the fluid particle which is at the position ${\bf x}$ at time $t$ 
(i.e. ${\bf X}(t'; {\bf x},t)$, in the notations of \S \ref{sec_LFDclosure}).
According to the transport equation for $\vct{\tau}(t)$ (Eq. \ref{eq:TranspEquTau}), 
the fluid particle is advected by the filtered velocity field. Thus, the position of the fluid particle 
at $t'$ is found by
backward particle tracking starting from end-time $t$ in the filtered velocity field. 
To perform
backward particle tracking, the filtered velocity and SGS stress
fields are calculated and stored at every $\Delta t=0.009$,
corresponding to $1/20$ of the Kolmogorov time scale. A Gaussian
filter is used with filter scale $\Delta=15\eta$, where $\eta$ is
the Kolmogorov length scale. In order to quantify isotropy as
function of $t-t'$, the ratio of off-diagonal to on-diagonal tensor
elements of the conditional averaged SGS stress at decreasing
previous time $t'$ is computed.

According to the derivation, the averaging must be conditioned on a 
particular value of the resolved velocity gradient $\overline{\textbf{A}}(t)$.
There are a large number of possibilities, since $\overline{\textbf{A}}(t)$ has 8 independent
elements. As representative of an important class of velocity gradient structure, we 
choose to consider regions where the $\overline{\textbf{A}}(t)$ is such that it has a large 
shear in one direction, whereas all other velocity gradient tensor elements are weak. 
We choose a particular shear direction, ``12'', and define  $\textbf{E}_{12}(t)$ to be a  ``high-12-shear" events that
occur at time $t$. These events are defined here as those points where 1) 
$\overline{A}_{12}(t) >
\overline{{A}}_{\rm rms}$, i.e. large and positive 12-shear, 2) 
$|\overline{A}_{ij}(t)|<
  \overline{{A}}_{\rm rms}$ for other off-diagonal components $(i,j) = 
(1,3)$ and $(i,j) =  (2,3)$, and 3) $|\overline{A}_{ij}(t)|< 
\overline{{A}}_{\rm rms}/\sqrt{2}$ for the diagonal elements $i=j$. The  
 gradient rms $\overline{A}_{\rm rms}$ is defined as  $\overline{{A}}_{\rm rms}\equiv\langle
\overline{A}_{ij}\overline{A}_{ij}\rangle^{1/2}$.
This definition allows a sufficiently large number of events to be counted and thus 
help in reaching statistical convergence. With this definition of a conditioning event,   we calculate
the isotropy factor $I(t-t')$ according to:
\be \label{eq:anisotropy}
I(t-t')\equiv -\frac{\langle \tau_{12} (t')
|\textbf{E}_{12}(t)\rangle}{(1/3)\langle \tau_{kk}(t')|
\textbf{E}_{12}(t)\rangle}.
\ee
 $I(t-t')$  monitors the isotropization of the SGS stress associated with
large ``12 shear events'', i.e. a particular anisotropic condition in the
large-scale velocity gradient tensor. 
Since the turbulence is statistically isotropic,  similar results are expected if the other two shear component
of $\tau_{ij}$, namely $\tau_{13}$ and $\tau_{23}$, had been chosen instead of $\tau_{12}$, under conditioning
based on events $\textbf{E}_{13}(t)$ or $\textbf{E}_{23}(t)$, respectively. 

Backward particle tracking starts from spatial
locations where the conditions in $\textbf{E}_{12}(t)$ are verified, at time $t$.
At each time $t'<t$, the particle locations are calculated from the
stored filtered velocity fields using a second-order Adam-Bashforth
scheme. The filtered velocity and SGS stresses at the particle
locations are interpolated from the stored fields using 6th order
Lagrangian interpolation. The conditional averages are then found by
averaging over all tracked particles. Statistical sampling is increased by
averaging over the trajectories starting from several different end
times $t$ and also over the other two $13$ and $23$ off-diagonal
elements (in both $ \tau_{ij} (t')$ and $\textbf{E}_{ij}(t)$).

\begin{figure}
\centering
\includegraphics[width=\linewidth]{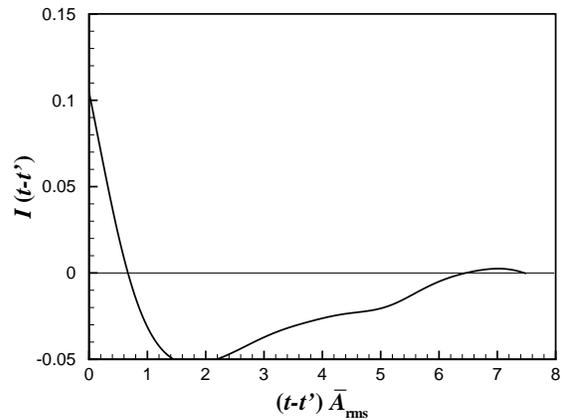}
\caption{\label{fig-i}Decay of the anisotropy factor $I$
(Eq. \ref{eq:anisotropy}) as function of normalized time lag $(t-t') \overline{{A}}_{\rm rms}$
measured in DNS of isotropic turbulence. }
\end{figure}
\begin{figure}
\centering
\includegraphics[width=\linewidth]{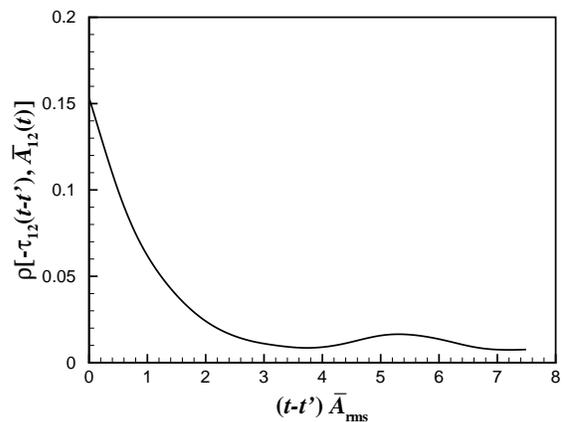}
\caption{\label{fig-rho}Decay of the correlation coefficient $\rho$ (see Eq.  \ref{eq:DefRho}) between
$\tau_{12}(t-t')$ and $-\ovl{A}_{12}(t)$ as function of normalized time lag $(t-t') \overline{{A}}_{\rm rms}$
measured in DNS of isotropic turbulence. }
\end{figure}

The resulting ratio $I(t-t')$  is plotted in Fig. \ref{fig-i} as a function of 
the normalized time lag $(t-t')
\overline{{A}}_{\rm rms}$. It is evident that the
conditional average of the SGS stresses becomes more isotropic as
the time lag increases and $I(t-t')$ crosses zero at
about 0.7 eddy turn-over times, namely  $t \sim 0.7 \overline{{A}}^{-1}_{\rm rms}$. 
Then there is negative undershoot to about negative half of the initial value, 
before it is relaxed to around zero (the isotropic value) at about
$t \sim 6 \overline{{A}}^{-1}_{\rm rms}$. The undershoot below zero is an interesting trend 
and understanding the physics of this behavior would be an interesting goal for future studies.

As additional evidence for the Lagrangian time decorrelation between stress
and large-scale velocity gradient, in Fig. \ref{fig-rho} we show
the correlation coefficient between  $\tau_{12}(t-t')$ and $-\ovl{A}_{12}(t)$, namely
\be \label{eq:DefRho}
\rho = -\frac{\langle \tau_{12}(t-t') \ovl{A}_{12}(t)\rangle}{\sqrt{\langle\tau_{12}(t-t')^2\rangle\langle
\ovl{A}_{12}(t)^2\rangle}}\mbox{ .}
\ee 
The correlation is near 15\% at zero time-lag (similar to the correlation coefficient between SGS stresses and strain-rate tensor often quoted in a-priori studies), but then decays to nearly zero 
at times around $t \sim 2 \overline{{A}}^{-1}_{\rm rms}$. Taken together, the DNS analyses thus provide evidence for the isotropy assumption on $\langle  \tau_{ij}(t') \vert \overline{{\bf A}}(t) \rangle$, as long as $t-t'  \gtrsim \tau_a$ with $\tau_a = \overline{{A}}^{-1}_{\rm rms}$. 
 
Note that due to the cost of storing the entire simulation for backward particle tracking, 
only moderate  Reynolds numbers were considered in the analysis. The forcing 
length scale has been estimated to be about 50 times  the Kolmogorov length scale, $\eta$, and the viscous 
effects begin to significantly damp the motions at scales of about  $10\eta$ and smaller. 
Therefore, using $\Delta=15\eta$, there may be some effects from the forcing and viscous scales on the results. 
However, the observed tendency towards isotropization is expected to become more, not less, prevalent
at higher Reynolds numbers.  We point out that opportunities for much more in-depth future analyses of such issues are provided
by the availability of a turbulence database at higher Reynolds number \cite{Lietal08} (although this database could not be used
for the present data analysis due to the fact that it does not yet contain sufficiently efficient means of filtering the data).

\section{Stress Tensor Model Based On The Recent Fluid Deformation Closure}
\label{sec_LFDclosure}

This alternative approach is based on relating the SGS stress tensor to 
small-scale velocity gradients. To begin, one may recall the multiscale expansion
\cite{Kra74,LiuMen94,Eyi05} in which, among others, the exact
subgrid stress (Eq. \ref{eq:DefSubgrid}) is written in terms of
$\textbf{u}^\delta$,  the velocity field coarse-grained at a scale
$\delta$, but still with $\delta<<\Delta$, i.e. containing significant
contributions from sub-grid scales. One may then define the approximated stress
tensor $\tau_{ij}^{\delta} = \overline{u_i^\delta
u_j^\delta}-\overline{u_i^\delta} \,\, \overline{u_j^\delta}$ and
naturally $\vct{\tau} = \lim_{\delta\rightarrow
0}\vct{\tau}^{\delta}$. 

Consistent with the Kolmogorov phenomenology, 
as argued formally in \cite{Eyi06a}, and also as used in various a-priori analyses of
experimental data (see e.g. \cite{LiuMen94}) the SGS stress is
relatively local in scale, stating that the leading terms entering
in its development are given by the coarse-grained velocity at the
resolution scale $\Delta$ and including also the next range of
length-scales between $\delta\approx \Delta/\beta$ and $\Delta$
(e.g. $\beta \sim 2$). As a consequence, one may use the
approximation $\tau_{ij}\approx \tau_{ij}^{\delta=\Delta/\beta}$.
Furthermore assuming that $\textbf{u}^{\delta=\Delta/\beta}$ is sufficiently
smooth over distances $\Delta$ (or using the `coherent subregion
approximation' \cite{Eyi06a}), a Taylor expansion of ${\bf
u}^{\delta}$ and evaluation of the filtering operation at scale
$\Delta$ in Eq. \ref{eq:DefSubgrid}  leads to
\be \tau_{ij}
\approx C_2 \Delta^2 ~{\partial {u}_i^{\delta}\over \partial
x_k}{\partial {u}_j^{\delta}\over \partial
x_k},~~~~~\delta={\Delta\over \beta} \label{eq-MSGstresss}\mbox{ .}
\ee One observes that similarity-type models such as the standard
nonlinear model \cite{ClaFer79,MenKat00} correspond to using the
gradient of the large-scale velocity field ($\delta=\Delta$ or
$\beta =1$). Nevertheless, it is the case $\beta > 1$ which is physically relevant
since the true SGS stress includes scales smaller than $\Delta$.
However, for $\beta > 1$, the expression \ref{eq-MSGstresss}
does not constitute a closure since then ${\bf u}^{\delta}$ contains
sub-grid motions that are not known at the LES filter scale
$\Delta$.

As in \cite{Gir94} and Chevillard \& Meneveau (2006 -- CM06 from here
on), a Lagrangian label position $\bf{X}$ is employed to encode
the time-history information. Using the two-time formulation of
\cite{Kra65}, the label positions ${\bf X}(t';{\bf x},t)$ satisfy
$d{\bf X}/dt' =\overline{\bf u}({\bf X}(t'),t')$ with ${\bf
X}(t)={\bf x}$. Thus ${\bf X}(t';{\bf x},t)$ represents the
position ${\bf X}$ at a prior time $t'$ of the fluid particle
which is at position ${\bf x}$ at time $t.$
Making the Eulerian-Lagrangian change of variables also used in CM06 leads to 
the following expression:

 \be \tau_{ij} = C_2 \Delta^2  ~
 {\partial X_p\over \partial x_k} {\partial X_q\over \partial x_k}
 {\partial {u}_i^{\delta}\over
\partial X_p}{\partial {u}_j^{\delta}\over
\partial X_q}.   \ee
All terms in this expression are strongly fluctuating variables. But, as 
in prior section, the most relevant information is retained by the 
conditional averaged expression. We propose the same
conditional averaging based on the time-history of the velocity gradient tensor along the
past fluid particle trajectory. Therefore, combining the conditional averaging and the
change of variables  one may write
 \be \label{eq:condav}
\tau^{(o)}_{ij}(t) = C_2 \Delta^2 \left\langle
 {\partial X_p\over \partial x_k} {\partial X_q\over \partial x_k}
 {\partial {u}_i^{\delta}\over
\partial X_p}{\partial {u}_j^{\delta}\over
\partial X_q}~\vert~ {\overline{\bf A}}(s);t' < s \leq t\right\rangle,   \ee
where the dependence of stress ${\vct \tau}^{(o)}(t)$ on current position ${\bf x}$ is
understood and not indicated to simplify the notation.
The Jacobian matrix $G_{ij}(t',t)=\partial X_i(t';{\bf x},t)/\partial x_j$ satisfies (see for instance \cite{TruNol92})
\be D_t {\bf G}(t',t) = - {\bf G}(t',t)\overline{{\bf A}}(t)
\label{G-eq} \mbox{ with } {\bf G}(t,t)={\bf I}\mbox{ ,} \ee where
{\bf I} is the identity matrix. Thus, $$\textbf{G}(t',t)= {\rm
Texp}^-\left[-\int_{t'}^t \overline{\textbf{A}}(s)\,ds\right]$$ is
expressed as an ``anti-time-ordered exponential'', with matrices
ordered from left to right for increasing times
\cite{DolFri79,Rugh96,FalGaw01}. The  only difference with the matrix function $\textbf{H}(t',t)$
of the preceding section (Eq. (\ref{eq:ODEforE})) is the sense of time-ordering.

Since the deformation gradient tensor $G_{pk}=\partial X_p / \partial
x_k$ is a deterministic function of the past velocity gradient
history, these tensors can be taken outside the conditional
averages in Eq. \ref{eq:condav}. So far one can thus write 
$$ \tau^{(o)}_{ij}(t) = C_2\Delta^2~
{\partial X_p\over
\partial x_k} {\partial X_q\over \partial x_k}~ Y_{ijpq}$$
$$ \mbox{with~ } Y_{ijpq} = \left\langle
 {\partial {u}_i^{\delta}\over
\partial X_p}{\partial {u}_j^{\delta}\over
\partial X_q}~\vert~ {\overline{\bf A}}(s);t' < s \leq t\right\rangle
$$ where ${\bf Y}$ is a 4th rank Lagrangian
gradient tensor. At this stage, it is now possible again to invoke Lagrangian isotropy, following the approach of CM06 and of the preceding section. It is assumed that the tensor  ${\bf Y}$ is isotropic due to loss of information caused by turbulent dispersion, past pressure
effects, etc. if $t-t'$ is long enough. Under the Lagrangian-isotropy closure assumption, one may write 

\be Y_{ijpq} = A'  \delta_{ij}\delta_{pq}+B'  \delta_{ip} \delta_{jq}+ C'  \delta_{iq}
 \delta_{jp}.
 \ee
While individual realizations of a small-scale
gradient tensor in turbulence are of course not isotropic, statistical moments such as the conditional average can be more justifiably approximated as isotropic.
The isotropy assumption states that the rate of change of
turbulent velocities $\delta {\bf u}^{\delta}({\bf x},t)$ (at the
present location and time $({\bf x},t)$), with respect to changes
in past locations of the fluid particles at time $t'$, is insensitive to orientation of $\delta {\bf X}$.
This appears to be a plausible postulate, if sufficient time has
elapsed, i.e. if $t-t'$ is sufficiently large for decorrelation to take place. In the preceding section we
used data from DNS to test the accuracy of such a de-correlation and 
`Lagrangian isotropy assumption'  in a closely related context (directly based on the stresses
rather than small-scale velocity gradient statistics). Still, it is important to
recognize that this step is introduced here as an `ad-hoc' closure
assumption and no claim is made that this is a formal step with
controlled errors.  

While the assumption of isotropy eliminates the dependence of 
$Y_{ijpq}$ upon ${\overline{\bf A}}$, the latter still affects the Jacobian matrix
$G_{ij}=\partial X_i/ \partial x_j$ that enters in the closure for the SGS stress.
Next, we focus attention only on the trace-free part of the modeled SGS stress tensor, i.e. we 
subtract the trace of the stress. And, noticing that the `right' Cauchy-Green tensor
$\partial_kX_p\partial_kX_q$ is symmetric, only the unknown $B' +C' $
enters in the resulting `quasi-optimal' model for the deviatoric part of the stress (superscript
$od$) model $\tau^{(od)}_{ij}$. Dimensionally, the parameter  $B' +C' $ has
units of inverse time-scale squared, and depends
upon the turbulence statistics down to scales $\delta$. For a fixed ratio $\Delta/\delta$,
and with both scales in the inertial range, for simplicity we  assume that the parameter $B'+C'$ follows, as in the prior section,  `Smagorinsky scaling', i.e. $B'+C' \approx  c ~ |\overline{\bf S}|^2$.
One thus obtains 

\be \label{eq:tauXXxx}
\tau^{(od)}_{ij}(t) = c_{\exp} ~\Delta^2|\overline{\bf S}|^2 ~
\left({\partial X_i\over
\partial x_k} {\partial X_j\over \partial x_k} -{1\over
3}{\partial X_m\over \partial x_k} {\partial X_m\over \partial
x_k}\delta_{ij}\right), \ee
where the parameter $c_{\exp}=C_2 \cdot c$, in a similar fashion as in the preceding section, is unknown and may be obtained by empirical knowledge, or by generalizing the dynamic model \cite{GerPio91}.

As a final step, the Recent Fluid Deformation (RFD) approximation
is used (CM06) in which  the time-varying velocity gradient
$\overline{\bf A}(s)$ between $t'$ and $t$ is approximated with
a constant value (e.g.)  equal to its  value at $t$ and denoted by $\overline{\bf A}$.
The initial condition for the fluid deformation (when the 
deformation gradient tensor is assumed to be the identity), is
prescribed at the time $t'<t$. The solution to Eq. \ref{G-eq} can
then be written as ${\bf G}(t',t) = e^{-(t-t')\overline{\bf
A}}$. Note that in this approximation,
$\textbf{H}(t,t')=\textbf{G}(t',t),$ since the sense of ordering of
the matrix products is no longer significant. 

The next step is to replace the solution for ${\bf G}(t',t)$ into Eq.
\ref{eq:tauXXxx}. And, as was done in the preceding section, 
to assume that a characteristic de-correlation time-scale $\tau_a$ has 
elapsed between the time where 
the initial isotropy assumption is justifiable and the current time when 
the stress closure is required. This  means replacing the initial time $t'$ with
$t-\tau_a$. Finally, 
the closure for the deviatoric part of the stress reads

 \be \vct{\tau}^{(od)}_{ij}  =  c_{\rm exp}
\Delta^2|\overline{\bf S}|^2 \left[ e^{-\tau_a\overline{\bf
A}}  e^{-\tau_a\overline{\bf A} ^\top } \right]^d  \label{eq:stressRFD}
\ee
where all quantities are evaluated at $({\bf x},t)$. 
It is immediately apparent that this closure is equivalent to the formal solution developed in the previous
chapter  (see Eq. (\ref{eq:stressTransportEq})).

\section{Expansions}
\label{sec_expansions}

In preceding sections it has been shown that a
matrix-exponential closure for the deviatoric part of the SGS stress tensor may be
written as in Eq. \ref{eq:stressRFD}. As a next step, the behavior of this closure is explored when
$\tau_a$ is small enough so that the norm of
$\tau_a \overline{\textbf{A}}$ is much smaller than unity. Then
$e^{-\tau_a \overline{\textbf{A}} } \approx {\bf I} -\tau_a \overline{\textbf{A}}
+(1/2)(\tau_a \overline{\textbf{A}} )^2+...$. Up to second order
one then obtains 
\begin{align} \label{eq:TaylorDevp1case} 
\vct{\tau}^{od} &\approx
c_{\rm exp}\Delta^2 |\overline{\textbf{S}}|^2 \Bigg(-2~
\tau_a ~\overline{\textbf{S}} \notag\\
& +~ \tau_a^2 ~\left[\overline{\textbf{A}}~
\overline{\textbf{A}}^\top +\frac{1}{2}\left(\overline{\textbf{A}}^2+
(\overline{\textbf{A}}^\top)^2\right)\right]^d~ +... \Bigg),
\end{align}
Crow, in Ref. \cite{Crow68}, derived essentially the same result (see his eq.(5.3))
but with unspecified coefficients obtained as moments of his memory
kernel  and an additional term proportional to the
material derivative $D_t\overline{\textbf{S}}.$
It is immediately apparent that if the time-scale $\tau_a$ is
chosen as $\tau_a = |\overline{\textbf{S}}|^{-1}$, then the first
term is the standard Smagorinsky model with $c_{\rm exp} = c_s^2$ 
(where $c_s$ is the Smagorinsky coefficient).
Furthermore, the second term, the term in the square parentheses, is of the
form of the `nonlinear model' \cite{ClaFer79,LiuMen94,MenKat00}
with a prefactor $c_{\rm exp} \Delta^2$. Two differences with the
standard  `nonlinear model' are apparent, however. The first is
that  if $c_{\rm exp} \sim c_s^2$, then as coefficient of the 
nonlinear term this is significantly smaller than the
coefficient for this term normally mentioned in the literature
(which ranges typically between 1/12 to 1/3). The second
difference is the presence of the additional term
$\left(\overline{\textbf{A}}^2+(\overline{\textbf{A}}^\top)^2\right)/2$.
To make connections with standard non-linear models used more
often in RANS (e.g. \cite{Spe91}), the velocity gradient is
decomposed into symmetric and antisymmetric parts,
$\overline{\textbf{A}}=\overline{\textbf{S}}+
\overline{\vct{\Omega}}$. The result is (again with $\tau_a =
|\overline{\textbf{S}}|^{-1}$)
\be\label{eq:TaylorDevp1caseSOmega}
\vct{\tau}^{od} \approx - 2 c_s^2\Delta^2|\overline{\textbf{S}} | \overline{\textbf{S}} +
c_s^2\Delta^2\left[\overline{\textbf{S}}^2
+\frac{1}{2}\left(\overline{\vct{\Omega}}\ \overline{\textbf{S}}-
\overline{\textbf{S}}\ \overline{\vct{\Omega}}\right)\right]^d\mbox{ .}
\ee 
It is interesting to note that the expansion including the term
$(\overline{\textbf{A}}^2+ (\overline{\textbf{A}}^\top)^2)/2$
cancels exactly the
$\overline{\vct{\Omega}}~\overline{\vct{\Omega}}$ part that is
included in the standard non-linear model
$\overline{\textbf{A}}~\overline{\textbf{A}}^\top$. For detailed
a-priori studies of the various decompositions of the velocity
gradient and non-linear terms see \cite{Hor2003}.

\section{Matrix exponential closure in LES of isotropic turbulence}
\label{sec_testsinLES}

The expansion introduced in the last section is formally valid only
for small values of the norm of $\tau_a \overline{\bf A}$. For more
realistic larger values, the expansion may be inaccurate and many
additional higher-order terms are needed. They can all be expressed
in terms of expansions into  integrity bases \cite{Pop75}, but it
is in general difficult to obtain the coefficients of the expansion.
Instead, it is proposed here to utilize the matrix exponential
directly in simulations. Since the exponential involves the full
velocity gradient tensor, it appears more natural to choose the
time-scale $\tau_a$ according to 
$\tau_a=\gamma (\overline{A}_{ij}\overline{A}_{ij}) ^{-1/2} \equiv \gamma |\overline{\textbf{A}}|^{-1}$
instead of using
$|\overline{\textbf{S}}|^{-1}$. The parameter $\gamma$
is an empirical coefficient of order unity.

As a first test, LES of forced isotropic turbulence is performed.
This flow is the simplest possible test-case and it is used here
simply to determine whether simulations using the matrix-exponential
based closure are numerically stable yielding realistic energy
spectra, and to ascertain the associated computational cost. 
The generalization to dynamic versions 
and tests in more complex flows will be left for future investigations. The
simulation uses the same pseudo-spectral method as was used in the
DNS outlined in \S \ref{sec_stressequation}, with same grid
resolution, forcing scheme and time step size. Dealiasing is
performed by zero-padding according to the two-third rule. The
viscosity of the fluid is $\nu=0.000137$. The subgrid-scale model
implemented is given by Eq. \ref{eq:stressRFD} and $c_{\rm exp}= (0.1)^2$ is
chosen (dynamic versions \cite{GerPio91} of this model to determined
$c_{\rm exp}$ can be developed in the future). 
To specify $\tau_a$,  the values $\gamma$=0.5, 1 and 2 are tested (a dynamic approach
of determining $\gamma$ could also be developed).  In
the pseudo-spectral   scheme, the modeled SGS stress is
evaluated in physical space and is made trace-free (this  only
affects the effective pressure, not the dynamics) before
computing its divergence in Fourier space.

\begin{figure}
\centering
\includegraphics[width=\linewidth]{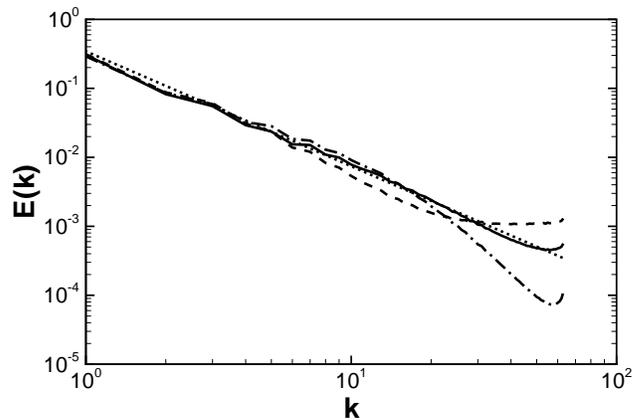}
\caption{\label{fig-ek} Radial kinetic energy spectra of
forced isotropic turbulence from LES  using the matrix-exponential
closure of Eq. \ref{eq:stressRFD} with $c_{\rm exp}=(0.1)^2$.
Solid line: $\gamma = 1$, dash-dotted
line: $\gamma=2$, and dashed line: $\gamma = 0.5$. Dotted line:
universal Kolmogorov spectrum  $E(k) = 1.6
\epsilon_f^{2/3}k^{-5/3}$.}
\end{figure}
\begin{figure}
\centering
\includegraphics[width=\linewidth]{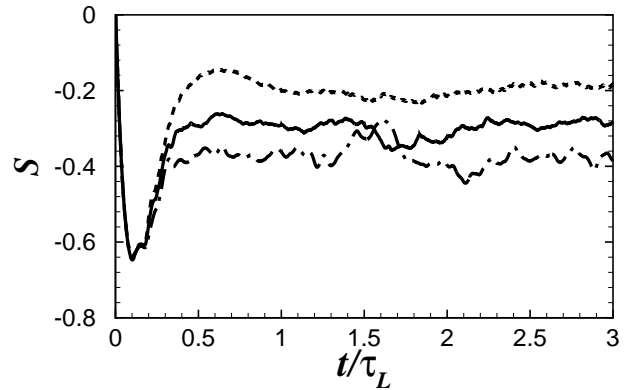}
\caption{\label{fig-s} Longitudinal derivative
skewness coefficient ${\cal S}$ as function of simulation time. Lines are the same 
as in Figure \ref{fig-ek}. $\tau_L \approx 6$ is the integral time scale.}
\end{figure}

The matrix exponentials are evaluated using truncated Taylor
expansion with scaling and squaring \cite{GolVan03}. Specifically,
we need to evaluate $\exp({\bf B})$, where ${\bf B}= -\gamma
\overline{\bf A}/|\overline{\bf A}|$. For a matrix $\bf C$ in
general, the $K$th order truncated Taylor expansion uses matrix
polynomial ${\bf T}_K({\bf C})=\sum_{n=0}^K {\bf C}^n/n!$ to
approximate $\exp({\bf C})$, incurring an error bounded by
$\parallel {\bf C}\parallel^{K+1}/\{[1-\parallel {\bf
C}\parallel/(K+2)](K+1)!\}$. The error decreases with the norm of
the matrix $\bf C$. Therefore, to evaluate $\exp({\bf B})$, we first
define ${\bf C}={\bf B}/2^j$, where the value of the integer $j$ is
chosen to ensure $\parallel {\bf C}
\parallel \leqslant 1/2$. $\exp({\bf C})$ is then approximated by ${\bf
T}_K({\bf C})$ and finally $\exp({\bf B})$ is given by $[{\bf
T}_K({\bf C})]^{2^j}$. The cost of calculating ${\bf T}_K({\bf C})$
is reduced by using Cayley-Hamilton theorem to express ${\bf C}^n$
($n>2$) in terms of ${\bf I}$, ${\bf C}$, ${\bf C}^2$, and the invariants of
${\bf C}$. Choosing $K=7$ , we obtain the following equation for
${\bf T}_7({\bf C})$  with an error smaller than $10^{-8}$:
\be
{\bf T}_7({\bf C})=C_0{\bf I}+C_1{\bf C}+C_2{\bf C}^2
\ee
where
\begin{eqnarray}
& &C_0=1-\frac{R_C}{3!} +\frac{Q_C R_C}{5!} - \frac{Q^2_C R_C}{7!},\notag\\
& &C_1 = 1- \frac{Q_C}{3!} - \frac{R_C}{4!} + \frac{Q_C^2}{5!} + \frac{2 Q_C R_C}{6!} + \frac{R_C^2-Q_C^3}{7!},\notag\\
& &C_2 = \frac{1}{2} - \frac{Q_C}{4!} - \frac{R_C}{5!} + \frac{Q_C^2}{6!} + \frac{2Q_CR_C}{7!}\notag \mbox{ .}
\end{eqnarray}
Here $Q_C=-{\rm Tr}({\bf C}^2)/2$ and $R_C=-{\rm Tr}({\bf C}^3)/3$
are the two non-zero invariants of $\bf C$ (note that ${\rm Tr}
({\bf C})=0$).  In terms of cost, the above algorithm uses about
$(1+j)N^3 + 5N^2+2N+37$ flops to
calculate $\exp({\bf B})$ when ${\bf B}$ is given, where $N$ is the
dimension of the matrix. In our tests $j=1+{\rm
floor}(\log_2\gamma)$, so $j=1$ when   $\gamma=1$ and the cost is
estimated at about 140 flops for each stress evaluation.
This can be compared with the
single matrix multiplication needed for the nonlinear model, which
is about $N^3 \sim 30$ flops. Overall with this closure, our code
took about twice as long to run as compared to using the mixed
model.

Simulations were initialized with random Fourier modes and evolved
until statistical steady state was obtained. No numerical
instabilities were observed for the three parameter cases
considered ($c_{\rm exp}=0.01$, $\gamma=$0.5, 1 and 2). In Figure
\ref{fig-ek}  the energy spectra obtained from the three
simulations as averaged in the time interval between one and three
large-eddy turnover times are shown. Figure \ref{fig-s} shows the
time-evolution of the derivative skewness coefficient ${\cal
S}=\langle (\partial_1 \overline{u}_1)^3\rangle/\langle
(\partial_1 \overline{u}_1)^2\rangle^{3/2}$. As can be seen, the
case $\gamma=1$ appears to yield physically meaningful results, which
can be compared with the well-known results of the Smagorinsky model
and the mixed model (see, for example, \cite{Kangetal03}). 
But, there is clear dependence on the parameter $\gamma$. The
skewness coefficient quickly drops to values near $-0.3$ for
$\gamma=1$ and $-0.36$ for $\gamma=2$. These are realistic values
for filtered turbulence \cite{Ceruttietal2000}. The skewness
values for $\gamma=0.5$, on the other hand, appear to be too close
to zero, consistent with some pile-up of the spectrum at high
wave-numbers.

\section{Discussion}

A new closure based on matrix exponentials and assumptions about
short-time Lagrangian dynamical evolution has been proposed.
Matrix exponentials as formal solution of the stress transport
equation provides interesting insights into the effects of the
production (gradient-stretching) term.   Historically, in the context of RANS modeling
using additional transport equations, the (closed) production term has justifiably not
been the focus of attention in the literature.  In LES, however, due to practical 
constraints `algebraic' closures are most often preferred. 
The present approach shows that the effects of production  in the
context of such closures  may be taken into account  directly based on an exact solution
of the stress transport equation. A central step in the present approach is to
use isotropy for the `upstream' initial condition. Evidence for such isotropization of
initial condition, given present large-scale velocity gradients, has been obtained using
a DNS database.   Implementation of the closure in LES of forced
isotropic turbulence yielded good results.  The computational cost
is significant, but it is not prohibitive. Since our code with
this model took about twice as long to run as with a traditional
algebraic closure, LES with this model at a resolution of $N^3$
has similar CPU cost as LES with a traditional  model run at a
resolution of $(2^{1/4}N)^3 \sim (1.2 N)^3$.

It is crucial to stress that the additional, 
more subtle physics of the remaining terms in Eq.\ref{eq:TranspEquTau} 
(pressure effects, turbulent diffusion, dissipation, etc..) 
are, in general, unlikely to be well-represented by the
simple assumption of `upstream' isotropy. In addition,  non-equilibrium
conditions in which ${\bf A}$ varies quickly along the particle
trajectory are not included in the closure as written in Eq.
\ref{eq:stressRFD}, in which the velocity gradient is assumed to
have remained constant over a time-scale $\tau_a$. To explore
non-equilibrium effects, the full time-ordered exponential
function must be used, although this would still leave out the
non-equilibrium effects of $\vct{\Phi}$.   
To compare the present approach to other closures  will require more
in-depth testing in more demanding, complex flows (e.g. where
effects of anisotropy, non-equilibrium, and pressure-strain
correlations are expected to be important).
 
It is also instructive to consider the case of 
two-dimensional (2D) turbulence. 
Nothing in the closure strategies pursued here limits their
application to space dimension three, at least nothing very
obvious. However, the expansions (Eqs.
\ref{eq:TaylorDevp1case},\ref{eq:TaylorDevp1caseSOmega}) show that
this is not likely to be a qualitatively good closure for space
dimension two, since one there expects an effective
``negative eddy-viscosity'' corresponding to inverse energy
cascade \cite{Eyi06b}. It is thus worth reflecting on some of
the reasons for the inaccuracy of the closures in 2D, since this
may help pinpoint potential shortcomings in 3D as well. First, it
is known that the 2D inverse cascade is less local than the 3D
forward cascade, with most of the flux coming from triadic
interactions for a scale-ratio $\beta=4\sim 8.$
\cite{Eyi06b,CheEck06}. However, the starting point of the RFD
closure,  Eq. \ref{eq-MSGstresss}, is not accurate for $\beta \gg
1.$ To get a qualitatively reasonable alternative at $\beta$
substantially larger than 1---which involves only first-order
gradients---one must instead use something like the ``Coherent
Subregions Approximation'' of \cite{Eyi06b}. On the other hand,
the starting point of the closure approximation in Section \ref{sec_stressequation}, 
the stress transport equation \ref{eq:TranspEquTau}, is exact in 2D
just as in 3D. The failure of the closure procedure in 2D is now
due, presumably, to the effects of the $\vct{\Phi}$ source-terms
in the transport equation. Indeed, those terms are expected to
contribute as an effective ``negative viscosity'', primarily due
to the pressure-Hessian rotating small-scale strain matrices
relative to the large-scale strain \cite{Eyi06b}.  
Note that, strictly speaking,
this is probably also true in 3D, so that the matrix-exponential
closures are likely to be overly dissipative in every dimension.
The main effect of the gradient stretching terms---which is a
tendency to forward cascade, or positive eddy-viscosity---is well
captured by the matrix-exponential closure in any dimension, but
the additional, more subtle physics of the remaining terms in Eq.
\ref{eq:TranspEquTau} are most likely not well-represented by the
simple assumption of isotropy. 

As has been cautioned several times in this paper, 
the simplified matrix-exponential  closure 
as written in Eq. \ref{eq:stressTransportEq}
employs the drastic approximation of entirely omitting the pressure-strain correlation and 
other `nonlinear scrambling' terms. But unlike eddy-viscosity based closure assumptions,
this expression can be derived directly from a relevant fluid dynamical equation, namely the  stress transport equation (with only the production term),  and using physically motivated and straightforward assumptions about  Lagrangian  decorrelation and  upstream isotropy. A similar result is obtained using an Eulerian-Lagrangian change of variables when the stress is expressed in terms of subgrid-scale velocity gradients. Perhaps it can be expected that casting this new light on the closure problem improves our understanding of this long-standing problem.

Finally, we remark that many transport equations for turbulence moments 
have a basic structure similar to Eq. \ref{eq:TranspEquTau}, including two 
production terms involving the velocity gradient and its transpose. 
Examples include higher-order
moments of velocity, the spectral tensor encountered in Rapid
Distortion Theory calculations, etc... The formal solution in
terms of matrix exponentials provides new  possibilities of
calculation and insights into the underlying physics.

\begin{acknowledgments}
We thank S.B. Pope and R. Rubinstein for useful comments. L.C. thanks the Keck
Foundation for financial support. The other authors thank the
National Science Foundation for financial support, and C.M. also
acknowledges partial support from the Office of Naval Research.
\end{acknowledgments}


\end{document}